\documentclass{aa}  

\usepackage{graphicx}
\usepackage{txfonts}
\begin{document}

  \title{ALMA CO(2-1) observations in the XUV disk of M83}
\titlerunning{ALMA observations in the M83 XUV disk}

   \author{Isadora C. Bicalho \inst{1}
          \and
          Francoise Combes\inst{1,2}%\fnmsep\thanks{Just to show the usag  of the elements in the author field}
          \and
          Monica Rubio \inst{3}
          \and
          Celia Verdugo \inst{4}
          \and
          Philippe Salome \inst{1}
          \fnmsep\
          }
\authorrunning{I. Bicalho et al.}

   \institute{Observatoire de Paris, LERMA, CNRS, PSL Univ., UPMC, Sorbonne Univ., F-75014, Paris, France\\
              \email{isadora.chaves@obspm.fr}
    \and
           College de France, 11 Place Marcelin Berthelot, 75005, Paris, France\\
              \email{francoise.combes@obspm.fr}
    \and
           Departamento de Astronomia, Universidad de Chile \\
             \email{mrubio@das.uchile.cl}
       \and
       	Joint ALMA Observatory, Santiago, Chile\\
		 \email{cverdugo@alma.cl}
             %\thanks{The university of heaven temporarily does not accept e-mails}     
             }
             % \date{Received April 19, 2017; accepted May 26, 2017}

 \abstract
   {The extended ultraviolet (XUV) disk galaxies are one of the most interesting objects 
studied in the last few years. The UV emission, revealed by GALEX, extends well beyond the optical
disk, after the drop of H$\alpha$ emission, the usual tracer of star formation. This shows that 
sporadic star formation can occur in a large fraction of the HI disk, at radii up to 3 or 4 times
the optical radius. In most galaxies, these regions are poor in stars and 
dominated by under-recycled gas, therefore bear some similarity to 
early stages of spiral galaxies and high-redshift galaxies. 
 One remarkable example is M83, a nearby galaxy with an extended UV disk reaching 2 times the 
optical radius. It offers the opportunity to search for the molecular gas and characterise 
the star formation in outer disk regions, traced by the UV emission. We obtained CO(2-1) observations with
 ALMA of a small region in a 1.5'$\times$ 3' rectangle located at $r_{gal}=7.85'$ over a bright UV region of M83. 
There is no CO detection, in spite of the abundance of HI gas, and the presence of young stars traced
by their HII regions. Our spatial resolution (17pc x 13pc) was perfectly fitted to detect Giant Molecular Clouds (GMC),
but none were detected.
The corresponding upper limits occur in an SFR region of the Kennicutt-Schmidt diagram
where dense molecular clouds are expected. Stacking our data over HI-rich regions, using 
the observed HI velocity, we obtain a tentative detection, corresponding to an
H$_2$-to-HI mass ratio of  $<$ 3 $\times$ 10$^{-2}$.
A possible explanation is that the expected molecular clouds are CO-dark, because of the strong UV radiation field. 
The latter preferentially dissociates CO with respect to H$_2$, due to the small size of the star forming clumps 
in the outer regions of galaxies. 
}


   \maketitle
%
%-------------------------------------------------------------------

\section{Introduction}

Over the last decades XUV disk galaxies have gained interest. Where \cite{GP05} shows the presence  of UV-bright complexes in the outermost of some galaxy disks. These are called  XUV disk galaxies, hosting UV emission well beyond their optical radii. UV-bright disks extending up to 3 to 4 times 
their optical radius (R$_{25}$) have been reported in about 30\% of spiral galaxies \citep{Thilker07,GP07a}. Their extended UV emission covers a significant fraction of the area detected in HI at 21 cm wavelength \citep{Bigiel10,Boissier2003,Z}.  Generally, the UV star-formation (SF) is related with this extending HI structure, e.g. shows evidence for metal enrichment \citep{GP07}.  Their far ultraviolet (FUV) and near ultraviolet (NUV) colours are generally consistent with young populations of O and B stars which probe a wider range of ages than H$_{\alpha}$ and at low SF levels the number of ionizing stars may be very small \citep{Dessauges, Boissier2003}.

Studying star formation beyond the optical radius allows us to address the condition of 
low-metallicity environments  \citep{Bigiel10} .
The study of SF in nearby galaxies involves the star formation rate (SFR), its surface density ($\Sigma_{SFR}$) 
and the gas surface density ($\Sigma_{gas}$) including both molecular and atomic gas.
The relation between these quantities is known as the Kennicutt- Schmidt (KS) relation  
({$\Sigma_{SFR} \propto \Sigma_{gas}^n$}, \citep{schmidt,kennicutt98,kennicutt12}). 
This relation describes how efficiently galaxies turn their gas into stars, in quantifying
 the star formation efficiency (SFE). The KS relation is almost linear when most of the gas is molecular, 
providing a constant gas consumption time-scale of about 3Gyr (e.g \cite{Bigiel11, saint11}). 
The SFE falls very quickly when the $\Sigma_{gas} < 10M_{\odot}/pc^2$ when the gas is mainly atomic. 
However, recent surveys of molecular gas at high resolution have the sensitivity to probe this 
relation at $\Sigma_{gas} \lesssim 3M_{\odot}/pc^2$. Studies of SF beyond the optical radius primarily 
concentrated on comparing different SF tracers. Until now studies of SFE in XUV disk was principally focused on atomic gas. 
These environments are not propitious to H$_2$ formation due to the low gas density and low
metallicity conditions, which are akin to galaxies in the early universe.

The confirmed occurrence of star formation in the outer disk of normal spirals has several important 
implications: it indicates the presence of molecular gas in the outskirts of spirals, possibly an efficient
phase transition from HI to H$_2$. It is one of the regions to study the unresolved issue of 
the atomic hydrogen gas origin: either HI is the main phase, that can be transformed to
H$_2$ to form stars, or reversely, it is the product of the star formation process, i.e. the result of the photodissociation of $H_2$ by the UV flux radiation emanated from newly formed stars \citep{Allen,smith00}. It also provides a simplified laboratory for determining the star formation 
threshold. It is also the place to investigate the star formation in quiescent and low-metallicity environments 
that may affect the SFE and the initial mass function.

 \cite{Crosthwaite02} studied the overall molecular gas morphology of M83 with CO(1-0) and 
CO(2-1) covering the 14'x14' optical disk.In this study they found that CO falls rapidly at 
radius of 5'=7kpc, and they proposed that this is also the decline in the total gas surface density,
even if the HI emission continues further out, but with a lower surface density. Molecular gas 
dominates inside 7kpc radius (80\% of the gas), while in total it is only 30\% of the gas. 
The CO(2-1)/CO(1-0) intensity ratio is $\sim$1.
At 7kpc, where the disk begins to warp, 
the ISM pressure might reach a threshold for the formation of molecular clouds.

  \cite{thilker05} with GALEX observations have modified this view: diffuse UV emission is 
detected beyond the bright star-forming disk, when H$\alpha$ and CO emission drop. 
This discovery made M83 the prototype of XUV disk galaxies. \cite{koda12} report deep 
Subaru H$\alpha$ observations of the extended ultraviolet disk of M83, and found some
weak emission, not seen by \cite{thilker05}. \cite{Dong08} with $SPITZER$ show that the SF 
has been an ongoing process in the extreme outer parts of M83 for at least 1Gyr.  
In a comparison between HI and FUV emission, \cite{Bigiel10} show that the most extended 
atomic gas observed in M83 will not be consumed by $in situ$ SF, and this might be due to the
low efficiency of the HI-to-H$_2$ phase transition there. However the present low SF might be
sufficient for chemical enrichment.
  A flat oxygen abundance gradient was obtained beyond $R_{25}$ by \cite{bresolin09}: they
find only a slight decline in abundance  
beyond this galoctocentric distance with 12+log(O/H) between 8.2 and 8.6. \cite{bresolin16} present 
a chemical evolution model to reproduce the radial abundance gradient of M83 in $R_{25}$,
and their model is able to quantify the metallicity of the gas, which is very close to that
of the stars.
 
 Star formation in low gas surface density, less than 10 M$_{\odot}$ pc$^{-2}$,
is not very well known.  The SFE in these regions
is very low and seems to be uncorrelated with $\Sigma_{gas}$: SFR has a much larger dynamical
 range than the local surface density \citep{Boissier2007}. CO observations in such environments 
are rare, due the weakness of the emission. A robust, quantitative picture of how the 
environment in the outer disks affects star formation is crucial to understand the origins 
of galaxy structure. \cite{Dessauges} were the first to study molecular SFE in an XUV disk, M63, 
where they detected CO(1-0) in 2 of 12 pointings using the IRAM-30m. The authors concluded 
that the molecular gas in those regions has low SFE compared to regions in the inner disc.  
There are only four galaxies with molecular detections in the outermost disks,
beyond $R_{25}$:  in addition to M63, NGC4414 \citep{Braine}, NGC 6946 \citep{Braine07},
M33 \citep{gratier10}. NGC4625 has been actively searched for CO emission but not 
detected \citep{watson}.
 In these papers, the SFE is defined as $\Sigma_{SFR}$/$\Sigma_{H_2}$, and we adopt this
definition here \citep{Dessauges}.

 In this paper, we present ALMA CO(2-1) data covering one region outside the optical disk of M83. 
We describe our ALMA observations in more details in Section \ref{sec:obs}, 
and our results in section 
\ref{sec:results}. Section \ref{sec:disc} presents the discussion about the SFR in low environments 
and physical reasons to explain the dearth of CO emission in the outer M83 disk.

\section{Observations}
\label{sec:obs}

\begin{figure*}
\centering
	\includegraphics[width=15cm]{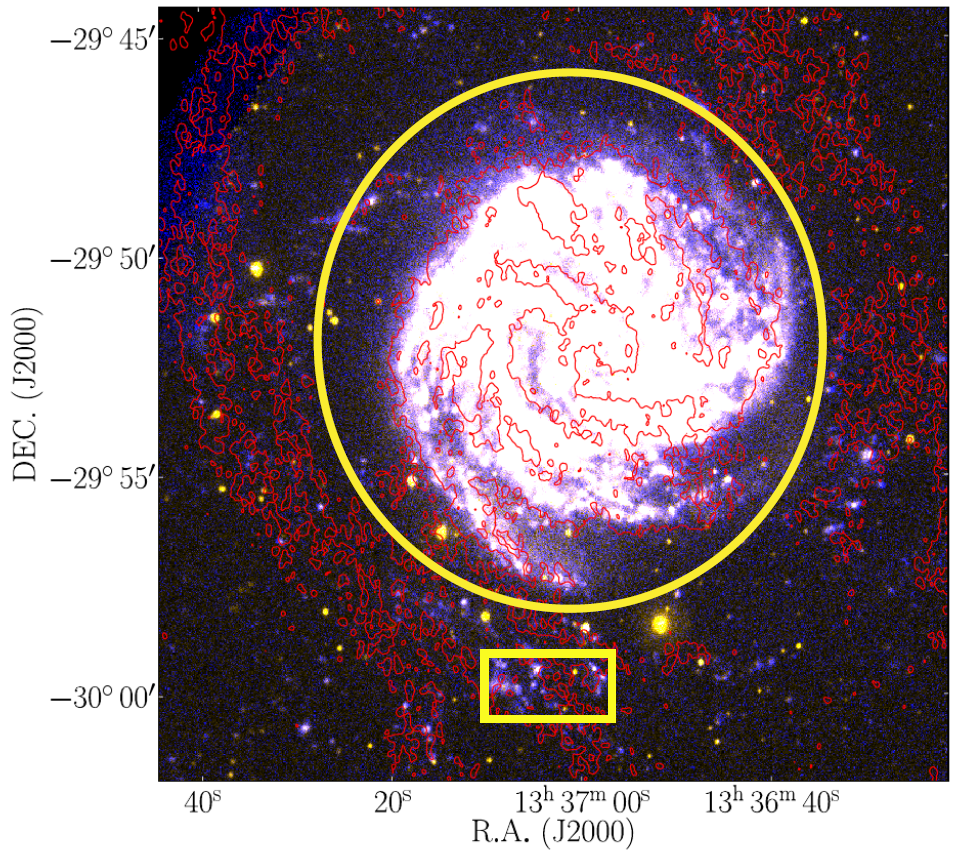}
    \caption{GALEX FUV image (background) of M83 with HI contours (red). The HI emission is from the THINGS 
survey. The yellow ellipse shows the optical radius ($r_{25} =6.09'= 8.64 kpc$) and the rectangle marks 
the region observed with ALMA and analyzed in this paper, located at $r_{gal}=7.85'= 11kpc$.}
    \label{fig:region_obs}
\end{figure*}

We observe CO(2-1) emission at 229.67 GHz (band 6) in M83 with ALMA during Cycle 2 (PI: Monica Rubio). 
The selected region is located at $13h 37m 03.6s ~ -29d 59' 47.6''$ enclosed 
in a $3'\times1.5'$ ($4kpc\times 2 kpc$)  rectangle located at $r_{gal} = 7.85' = 11kpc$ from M83 center. 
We used the selection criteria corresponding to FUV/NUV GALEX images \citep{GP07}, using the peaks 
of UV emission, as well as the correlation with the HI emission from the THINGS survey \citep{things}. 
By choosing these peaks of emission we focused on the outer parts of the UV disks, beyond the $r_{25}$ 
optical radius, where we are interested in detecting the molecular gas (see figure \ref{fig:region_obs}). 

\begin{table}
	\centering
	\caption{M83 observations}
	\label{tab:caract}
	\begin{tabular}{lclcl} % four columns, alignment for each
		\hline \hline& Values \\  \hline
		R.A (center) & $13h 37m 03.6s $ \\
		Decl. (center) & $-29d 59' 47.6''$\\
		Distance (Mpc)  & 4.8\\
		R$_{observed}$  & 7.85' = 11kpc\\
		Synthetized Beam & 0.75"x0.56" = 17x13pc\\ \hline
	\end{tabular}
\end{table}

\begin{figure*}
\centering
\includegraphics[width=15cm]{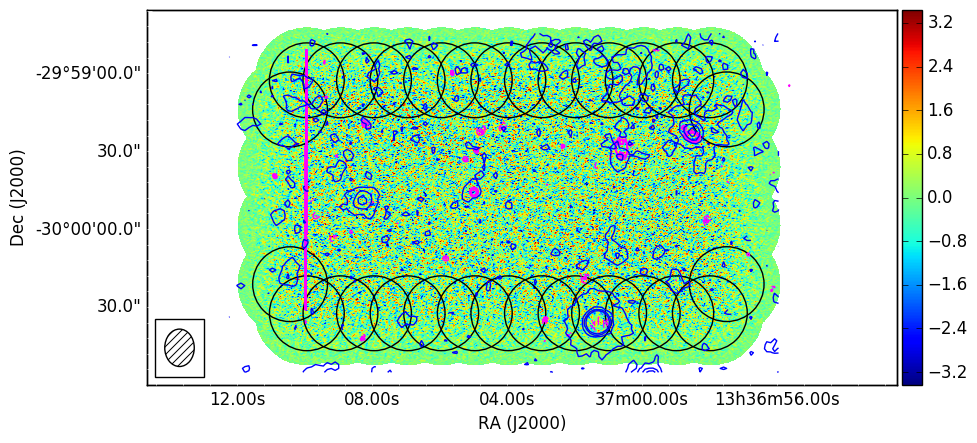}
    \caption{Scheme of the ALMA mosaic of our observations. The black circles of 27" diameter 
refer to the position of the 121 pointings used to map the CO(2-1) emission. The background image is 
the CO integrated emission (zero moment of the data cube). The contours are H$\alpha$ (magenta color) 
and FIR 24$\mu$m (black). }
    \label{fig:mosaic_point}
\end{figure*}

These selection criteria were used for the M63 XUV disk (NGC 5055), and led to a CO detection 
\citep{Dessauges} far outside the $r_{25}$ limit, while \cite{S11} only obtained an upper limit 
at 300" from the galactic center.  The selected region of M83 was observed during ~1hour in March 2014, 
in very good weather conditions (pwv~1.3mm). The 12m array was used with 34 antennas and a maximum 
baseline of 558.2 m. The map was done with a mosaic of 121 pointings separated by 12.9" 
(figure \ref{fig:mosaic_point}), and with an integration time of 10.8 sec per pointing.   

The data were calibrated using the CASA reduction package. Approximately 36\% of the data was 
flagged which was current at this epoch for 12m Array data. 
We produced a CO(2-1) data cube for each pointing with natural weighting, and for a velocity range 
from 15 to 1015 km/s, a channel spacing of 2.5 km/s and an rms of 10.3 mJy per channel. 
The calibrated 121 uv-tables were subsequently exported to GILDAS where the cleaning and the 
cube analysis were performed.  

There is no continuum detection, except
 one weak ($<$2mJy) continuum (point) source, most probably a background source,
at RA= 13h 37m 00.79s, DEC= -30$^\circ$ 00' 10.8''.
Using natural weighting, the synthesized beamsize of the
continuum map was 0.78"x0.60", with rms$=$0.19 mJy.

 As for the CO line, We can estimate
the upper limits found in each beam, assuming a profile width of 15 km/s FWHM. 
The rms noise level is 4mJy in 15km/s channels
for each beam of 0.75"x0.56".
 The 3$\sigma$ upper limit of the integrated
emission is therefore 0.180 Jy km/s. This corresponds to
L'CO(2-1) = 2.5 10$^3$  K km/s pc$^2$. Assuming an intensity ratio of
$I_{21}$/$I_{10}$ 0.7 (e.g. \cite{BraineCombes}), and a standard CO-to-H$_2$
conversion factor $\alpha = 4.36 M_\odot/(K km/s pc^2)$, this corresponds
to $M(H_2) = 1.5  10^4 M_\odot$. 
This mass scale
is much lower than a Giant Molecular Cloud (GMC) in the beam
of 17 x 13pc.

Another way to see this limit is to consider that our 10.3mJy/beam
noise level in 2.5km/s channels corresponds to a brightness temperature of 0.55K.
Therefore clouds of 1.6K should have been detected at 3$\sigma$. In the outer parts of the Milky Way,
as far as R=20kpc from the Galactic center, molecular clouds up to 10K (in CO(1-0)) and 5K (in CO(2-1))
have been observed, at similar widths \citep[e.g.][]{Digel1994, Sun2015, Sun2017}. They would have been
detected quite easily in our survey.

\section{Results}

\label{sec:results}

We searched in an automatic way for CO (2-1) emission with an SNR of more than 5 $\sigma$ in each 
of the 121 cubes, with the "detection assessment" procedure in GILDAS. 
We found 14 possibilities of emission, falling in the HI range
of velocities (500-700km/s) in this region of the galaxy,  out of the 121 cubes. 
However, there is no spatial coincidence between these hints of emission and any 
of the other tracers. Also the width of the profiles are in general too large.
Therefore evidence for the CO emission from molecular clouds is very weak.

\begin{figure}
	\includegraphics[width=\columnwidth]{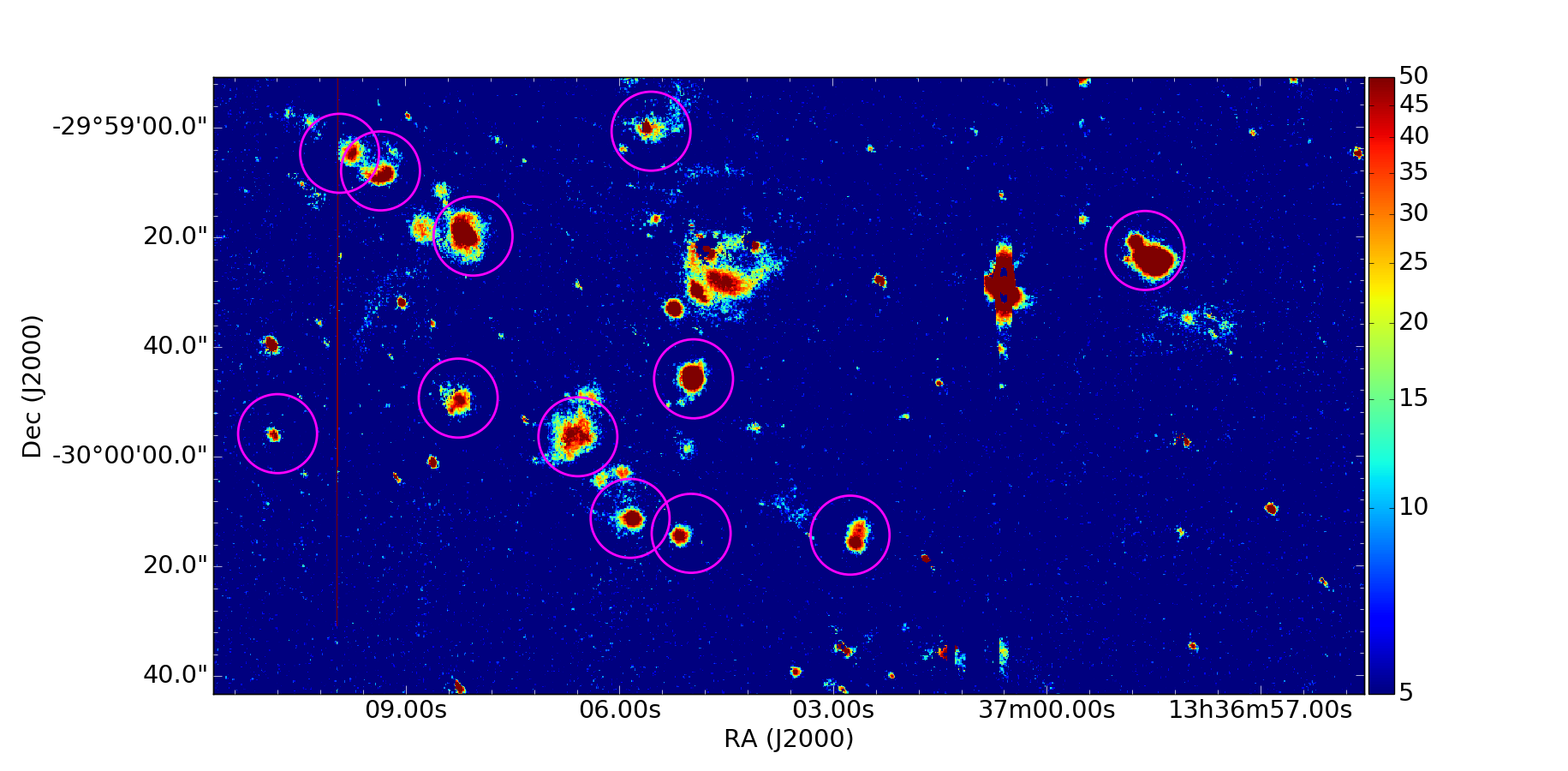}
    \caption{The $H_{\alpha}$ image of the region is shown, from Subaru \citep{koda12}.
The magenta circles show the positions where CO was searched for.}
    \label{fig: halphapos}
\end{figure}

Reversely, we look for the most prominent clumps from other tracers, in particular H$\alpha$.
Using the observations from the Subaru telescope \citep{koda12} we found 13 regions of star formation,
and tried to search for CO emission there.  These regions are shown in the figure \ref{fig: halphapos}, 
where the circles scheme the spatial regions correspondant to our choice. After looking to the ALMA data,
none of the hints found have more than 3$\sigma$ signal.

\subsection{Matched Filter Technique}

\begin{figure} 
	\includegraphics[width=\hsize]{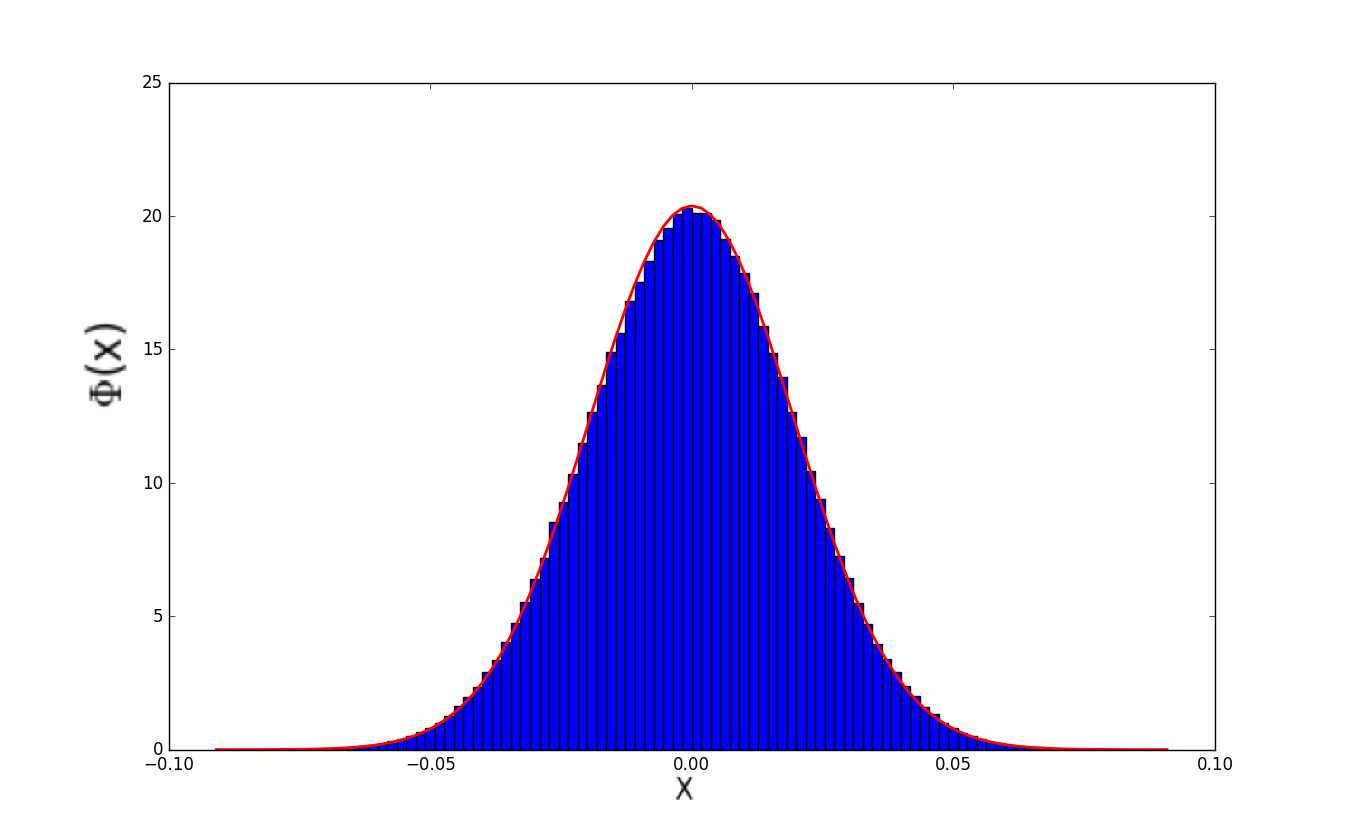}
    \caption{Histogram of the pixels values from the ALMA data cube (our observations). The blue histogram comes 
from the pixel values and the red curve is a Gaussian fit. We see that the noise is a perfect gaussian 
which does not show any indication of some bump that could be related to the reduction and deconvolution
procedures.}
    \label{fig:hist}
\end{figure}

The reliable detection of weak signals is an issue in astronomical data.
The data analysis process can lead us to underestimate the probability of false detection, 
that is why it is necessary to estimate it. 
A technique that is expected to have among the best probability of true detection 
is the Matched Filter (MF) technique.
The goal of the filter is to maximise the detectability of  signal of known structure inside a 
random noise Gaussian \citep{Vio16}.
To better comprehend the quality of our observations (e.g. to find out 
whether our measurements are biased) we 
first apply a simple technique. 
Here we want to check the assumption that the probability 
density function (PDF) of the noise peaks  is close to a Gaussian.  In \cite{Vio16}, 
they show from a zero-mean map that when the positions and number of sources are unknown
the matched filter could  underestimate the probability of false detections. 
We use here the simplest technique, 
where we plotted the pixels values from our moment 0 data. The values are plotted in 
figure \ref{fig:hist}. The plot shows no obvious irregularity or departure from a 
Gaussian, revealing no problem with the cleaning or reduction.  

 Since we know that the signal should correlate to the other ISM tracer, the HI-21cm emission, which 
gives us the spectral region to find the line, we now apply this filter to the data,through a stacking technique in the next section.

\subsection{ Stacking of CO spectra, according to the HI velocity}

\begin{figure*}
\centering
	\includegraphics[width=6cm]{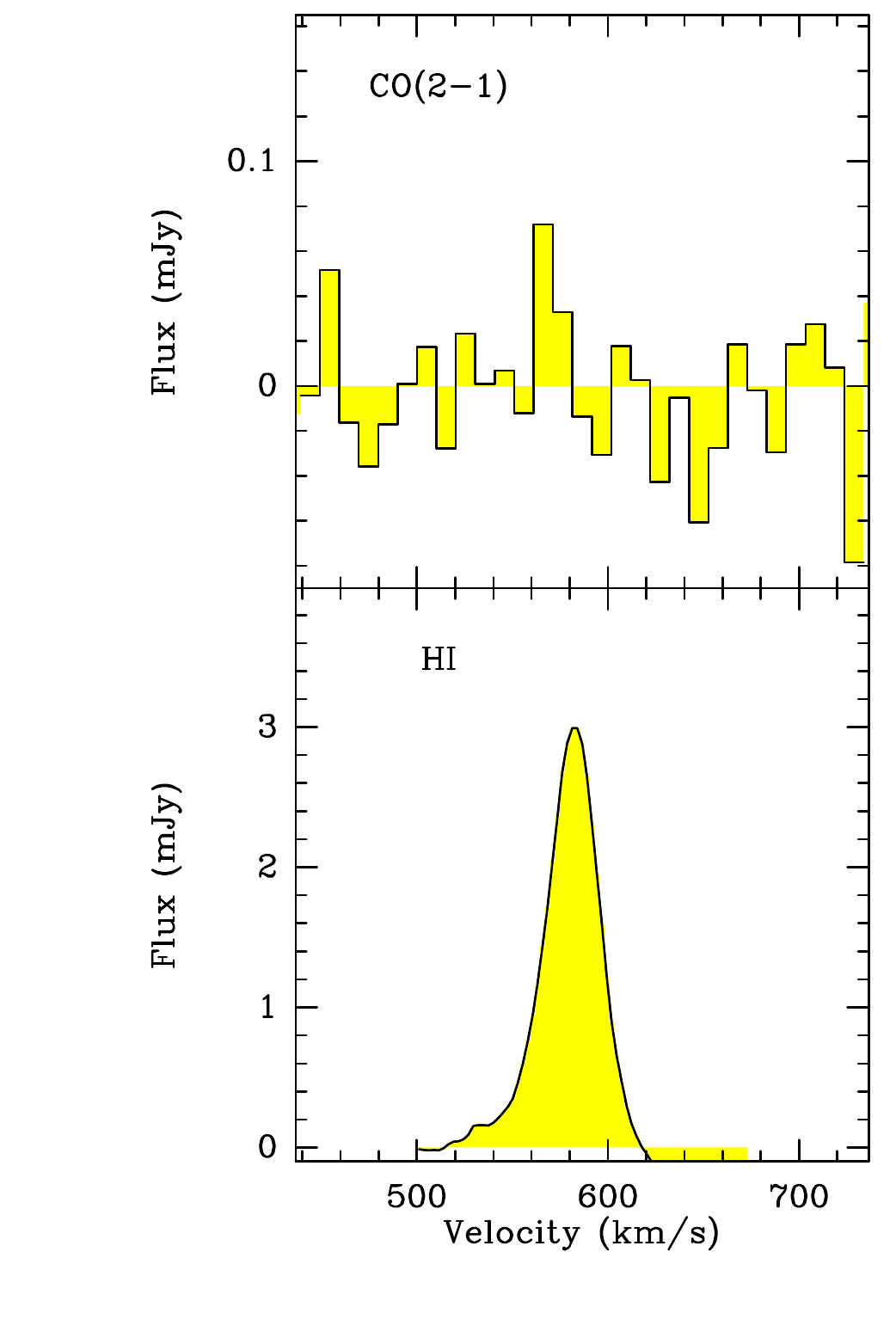}
	\includegraphics[width=10cm]{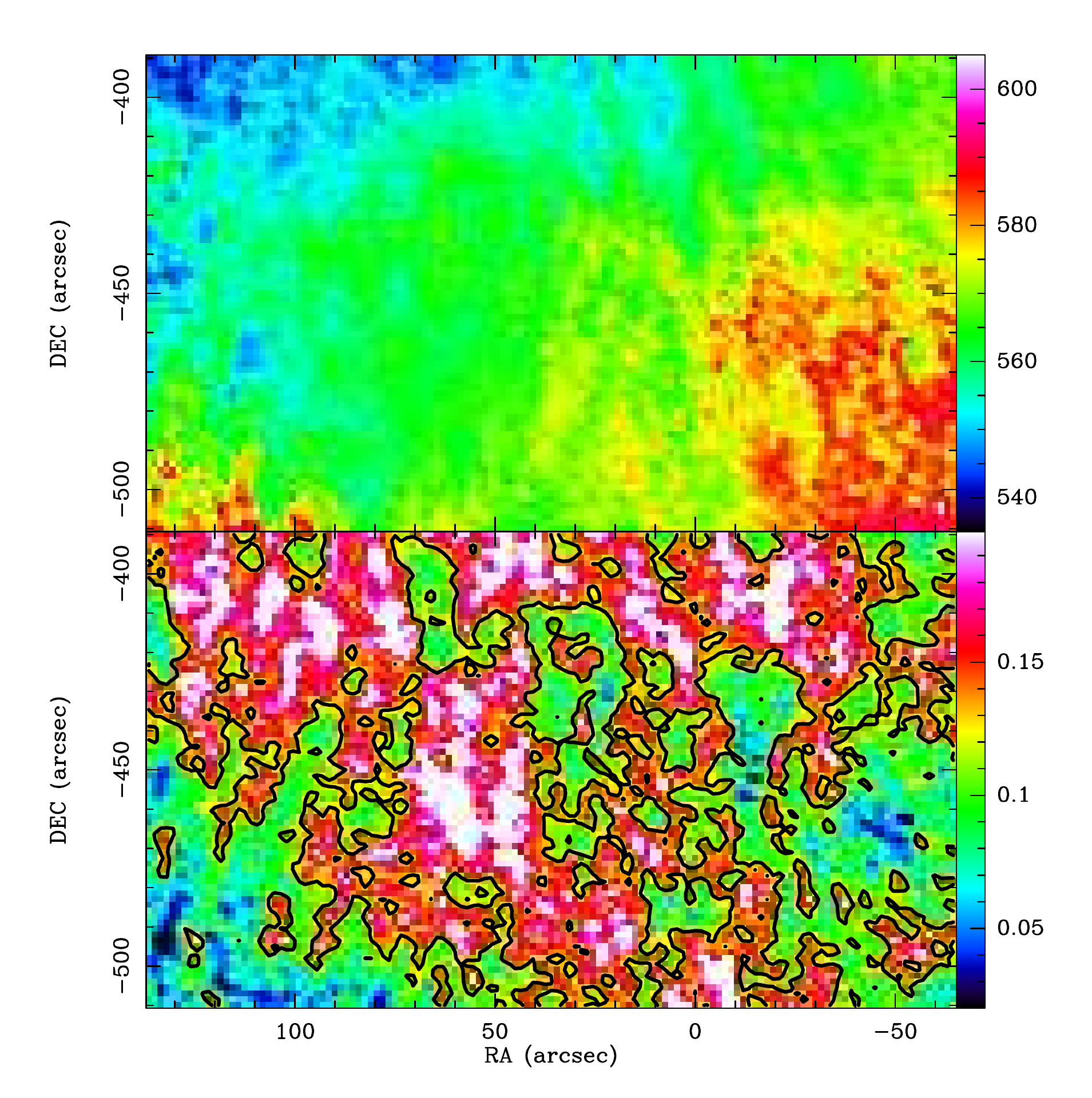}
    \caption{\textit{Left:}Result of the stacking procedure for the CO(2-1) map (top)
and the HI map (bottom). Only spectra with a significant HI detection have
been considered (see text), and they have all been shifted to 580km/s central
velocity. \textit{Right}: HI (THINGS) integrated intensity map (Jy$\times$km/s, bottom) in the 
region of the mosaic observed by ALMA, and the corresponding average velocity (top) in km/s. 
The spatial coordinates are in arcsec offset from the galaxy center 
(13h37m00.9s, -29$^\circ$51'56''). The HI beam is 15.2" x 11.4" (VLA NA map from
THINGS). The black contour corresponds to NHI = 10$^{21}$ cm$^{-2}$. }
    \label{fig:FC}
\end{figure*}

The CO emission is normally associated  with HI peaks, because the host molecular clouds 
probably form in regions of relatively high HI column density, 
in absence of spiral density waves \cite{Gardan} . 
Therefore, to exploit the full mosaic of data from ALMA, 
we must stack all individual spectra where emission
is expected, at the HI expected velocity. Averaging all pixels of
the map has not given any signal; this might be due to velocity
dilution, since there is a significant velocity gradient, of $\sim$
100km/s over the corresponding HI map (cf Figure \ref{fig:FC} at right).
 For this CO averaged spectrum, we find an rms of 0.017 mJy in 14km/s
channels.
 We therefore computed the first moment of the HI cube, to
have an average velocity at each 1.5$\arcsec$ x 1.5$\arcsec$ pixel of the map.
 We smoothed the CO(2-1) map at the 1.5$\arcsec$ resolution, and 
for each beam, we shifted the spectra of the right velocity
amount to have all of them centered at the expected velocity,
known from the HI spectrum.
Then we averaged all spectra, where the HI integrated intensity
was above a certain threshold, which was 50 mJy/beam $\times $  km/s. 
The same stacking was done on the HI cube, which gives the 
result plotted in Figure \ref{fig:FC} at left. The resulting 
HI stacked spectrum has a FWHM of 33 km/s. No baseline was
subtracted to the stacked spectra. The stacked CO(2-1) spectrum
shows a hint of emission (at 3.5$\sigma$) with a FWHM of 14 km/s.
When properly reduced, the average integrated flux is 0.85 mJy km/s.
Adopting the conversion factor described in section \ref{sec:obs}, 
this low value corresponds to an average mass of 65 M$_\odot$ of molecular
gas per beam, but spread over 3 x10$^4$ beams, therefore would represent
a total mass of 2 x 10$^6$ M$_\odot$ over the whole mapped region,
of 4 x 2 kpc.  The same computation can be done on the averaged
stacked HI spectrum. With a flux of 0.1 Jy/beam km/s, and a beam of
15.2\arcsec x 11.4\arcsec, this gives a total HI mass of 7.0 x 10$^7$ M$_\odot$ over the
same region. The mass ratio between the molecular and atomic gas is then
M(H$_2$) /M(HI) = 3 $\times$ 10$^{-2}$ or below.

 Let us emphasize that the stacking technique is not used here to 
smooth out possible extended signal on scales that, anyhow, would be filtered
out by the interferometer. On the contrary, we here aim at conserving our spatial resolution
of 17pc x 13pc fitted to GMC scale, in order to avoid dilution of 
the cloud emission. The stacking, as usual, is averaging out several
realizations of possible cloud emission.

\section{Discussion}

\label{sec:disc}

The extended ultraviolet disks, present in 10\% of nearby galaxies, offer the opportunity to 
study the interstellar medium and star formation in extreme conditions with low average gas 
density and surprisingly abundant star formation. M83 was one of the first XUV detection in 
its outskirt regions and it is the prototype for these type of galaxies. 
However, no highly significant 
CO emission was detected in Cycle 2 ALMA maps. To analyse the impact of our observations in the 
context of star formation in  outer XUV disks, we investigate the behaviour in the Kennicutt-Schmidt 
diagram of the tentative CO detections, in plotting the equivalent molecular gas surface density,
and star formation surface density.

\subsection{Star formation diagram}

The SFR surface density can be determined from a combination
of FUV and 24$\mu$m fluxes, with time-scale 10-100 Myr, 
 using the calibration from \citep{leroy2008}: 

\begin{equation}
\begin{aligned}
\Sigma_{SFR} (M_{\odot}yr^{-1}kpc^{-2}) ={} 
& 8.1 \times 10^{-2} F_{FUV} (MJy sr^{-1}) \\
& +3.2 \times 10^{-3} F_{24 \mu m} (MJy sr^{-1}) \
\end{aligned}
        \label{eq:sfr-fuv}
\end{equation}

\noindent  Within an extended HII region, the very recent SFR (3-10 Myr time-scale) can be
determined from the H$\alpha$ luminosity by \citep{kennicutt12}:

\begin{equation}
    SFR (M_{\odot}yr^{-1}) = 5.37 \times 10^{-42}  L(H\alpha) (erg s^{-1})
        \label{eq:sfr-alpha}
\end{equation}

\noindent The H$\alpha$ luminosity was corrected for internal extinction, 
through $ L(H\alpha)_{corr} = L(H\alpha)_{obs}+0.020 L(24\mu m)$, according
to \cite{kennicutt12}.
Over an aperture of 200 pc size around the two strongest HII regions,
we find an SFR of 3 $\times$ 10$^{-4}$ and  3.6 $\times$ 10$^{-4}$ M$_{\odot}$yr$^{-1}$.
 These values are comparable to what is obtained from equation \ref{eq:sfr-fuv}.
 The upper limits of CO emission are also averaged over
the same area (200 pc size), to compute the corresponding surface densities.
\cite{Schruba2010} have studied the dependence of the molecular
depletion time in M33 on the scale considered, and they conclude that 
$\sim$ 300pc is the limiting scale below which the SF law as a function of gas
surface density is likely to break. We consider a very
similar scale, where the SF law should be relevant.

We used the following equation to calculate the molecular hydrogen surface density: 

\begin{equation}
    \Sigma_{H_2} (M_{\odot}pc^{-2}) = 4.2  I_{1-0} (Kkm s^{-1})
	\label{eq:mol_h}
\end{equation}

\noindent where $I_{1-0}$  is the CO(1-0) line intensity in $K kms^{-1}$, and we assume that the 
CO(2-1) to CO(1-0) intensity ratio is $R_{21}=I_{21}/I_{10}= 0.7$. 
With the spatial resolution of our observations, the flux scale is 0.018 Jy per K. 
The standard CO-to-H$_2$ conversion ratio of $X_{CO}$ = 2 10$^{20}$ cm$^{-2}$ /(K km s$^{-1}$)
is adopted, and the number in equation \ref{eq:mol_h} includes  helium correction. 

\begin{figure}
\includegraphics[width=\columnwidth]{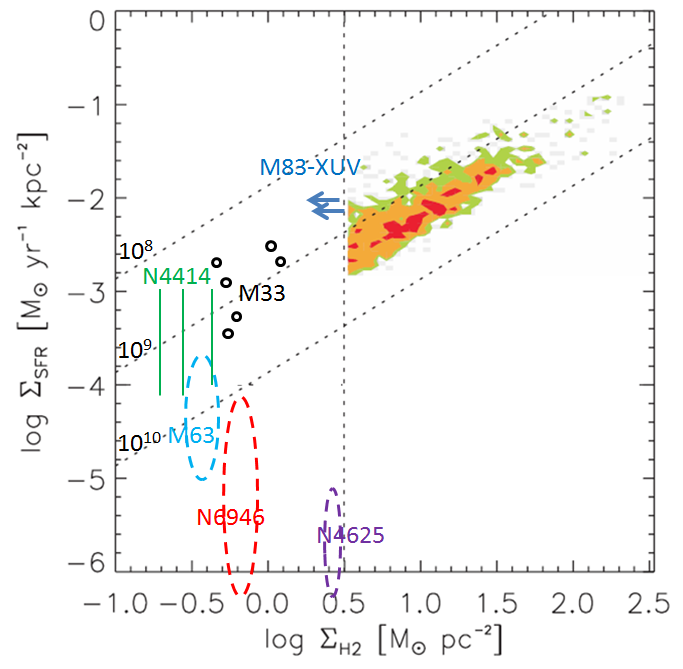}
    \caption{Kennicutt-Schmidt diagram relating the SFR surface density to the molecular
gas surface density, adapted from \cite{Verdugo15} and \cite{Bigiel08}. 
Dashed ovals represent the data from the outer parts of XUV disk galaxies:
NGC4625 and NGC6946 from \cite{watson}, and M63 (NGC5055)
from \cite{Dessauges}, taking only the molecular gas into account in all of them.
The 3 green vertical bars are from NGC4414 
\citep{Braine}, and the black circles for M33 \citep{gratier10}.
The dashed vertical line corresponds to 3 M$_\odot$/pc$^2$, the sensitivity limit 
of the CO data in \cite{Bigiel08}.
Dashed inclined lines correspond to depletion times
of 10$^8$, 10$^9$ and 10$^{10}$ years to consume all the gas at the present SFR.
 The horizontal upper limits correspond to our CO(2-1) results on the M83 XUV disk of $\Sigma_{H_2}$ ,
in two of the main SFR regions, of sizes 200 pc.
 } 
\label{fig:ks-h2}
\end{figure}

 We plot the Kennicutt-Schmidt relation in figure \ref{fig:ks-h2}, where we
compare all molecular data obtained in outer disks up to now,
with the large sample of nearby galaxies from \cite{Bigiel08}. The total gas surface density 
$\Sigma_{H_2}+\Sigma_{HI}$ is completely dominated by molecular gas above
9 M$_\odot$/pc$^2$, and the SFR relation can then be considered as linear with the
gas surface density.

For M83, the regions used for this plot are the most prominent
star forming regions in the H$\alpha$ map. We have smoothed the molecular gas
surface density over the extent of the H$\alpha$ region size, about 200 pc,
which leads to lower upper limits.
In those regions, the average SFR surface density is high enough that we could expect to find only
molecular gas. We then plot only the upper limit on H$_2$ surface density obtained by averaging over the
200 pc-sized region. For all other compared galaxies, only the molecular component is taken
into account.
It can be seen that the time to consume the molecular gas in the outer
regions of M83 is smaller (t$_{dep} <$ 3 10$^8$ yr) then the depletion 
time in nearby galaxies (t$_{dep} =$ 3 10$^9$ yr).
It is not relevant to represent in this diagram the average value obtained through the stacking
over the whole region of 4kpc x 2kpc, since the surface densities are then diluted
to a region much larger than the usual star forming regions, used in this diagram 
(about 200-500 pc in size).
 It is indeed over such scales that a star forming region can be defined. At lower scales, 
the star formation tracers, such as H$\alpha$ or FUV, are not expected to be correlated
to the molecular clouds of their birth, since newly born stars progressively drift away
 at a different velocity than the 
dissipative gas.  For the other galaxies, the gas and SFR surface densities concern also such scales, 
like M63 \citep{Dessauges} corresponding to the IRAM-30m telescope beam of 0.5 and 1kpc 
in CO(2-1) and CO(1-0) respectively.

\subsection{The dearth of CO emission in the outer disc of M83}

One main issue to explain the absence of CO emission in galaxies is the metallicity Z of the gas. 
It is now well established that CO emission is frequently very faint or nonexistent in low-metallicity galaxies,
for example in gas-rich irregular galaxies \cite{Elmegreen,taco_young,taylor}. 
Both observations and theory indicate that the CO-to-H$_2$ conversion factor, $X_{CO}$, increases at low metallicity
 below $12+log(O/H)=8.2-8.4$  \cite{Bolatto13}. The dependence of this relation is non-linear with Z, and may be in Z$^{-2}$ or steeper
at low Z, because of UV photo-dissociation in the absence of dust, in addition to the under-abundance of CO.

However, in the M83 field observed with ALMA, there is only little abundance deficiency,
the average metallicity of the H$_{II}$ regions is $12+log(O/H)=8.4$ \cite{bresolin09}, i.e.
1/2 solar. With a slight 
change of $X_{CO}$ for a higher value than standard, we should still be able to detect molecular clouds at more than 8$\sigma$. 
This means that the scatter in the $X_{CO}$-metallicity relation could not explain the dearth of CO emission. 

The formation of the CO molecule occurs from OH via ion-neutral reactions that form HCO+, 
which after its dissociation 
forms CO. Thus, the CO formation rate depends on the abundance of OH, itself related to the rate of 
destruction by UV photons. The destruction of the CO molecule is mainly by photodissociation. 
 In diffuse environments, the molecular clouds are more isolated, and less shielded.
A schematic view of a molecular cloud includes a dense core, with CO emission, and a surface
where CO is photodissociated, and emitting essentially in C+ and C lines
(see figure 8 in \cite{Bolatto13}). 
Because H$_2$ and CO have similar dissociation processes in their line-transitions, H$_2$ can partially shield the 
CO molecule, but with large CO column densities, the CO molecule is self-shielding. At low CO
column densities, and in particular in the outer parts of molecular clouds, the CO molecule is absent,
and the CO/H$_2$ abundance ratio drops. All the carbon is found in C and C+, and this
gas is called CO-dark molecular gas.

Another possible scenario which could explain the lack of CO emission is that the photo-dissociated region
  is fragmented in smaller clouds. Each one has not enough depth to have CO emission, 
and is dominated by C+ emission. 
In outer disks, the small star forming clouds will emit much less together,
for the same amount of gas,
than the larger clouds present within the optical disk. The same scenario was invoked in \cite{Lamarche} where they failed to detect CO emission towards the starbursting radio source 3C368 
at z=1.13 using ALMA data. They discuss the possibility that they observe
far-infrared fine-structure oxygen lines in star-forming
 gas clouds before they have had the chance to form an appreciable amount of CO.

\cite{watson} did not have a precise reason for not detecting CO in a 
young star-forming region of NGC4625, 
while \cite{Braine07}, \cite{Braine} and \cite{Dessauges} could detect weak
CO emission in the outer parts of NGC 6946, NGC 4414 and M63, respectively. However, in M63 
CO is detected only in two out of 12 XUV-disc regions. \cite{watson} conclude that,
even if one explanation for the dearth of CO emission in
lower-mass galaxies can still be the low metallicity, the issue remains for
the outskirts of massive galaxies. Deeper observations are needed to disentangle
the various proposed scenarios.

 The gas metallicity is only half solar in our mapped ALMA field, which
could already reduce the size of the CO clouds with respect to H$_2$ clouds.
The dearth of CO emission could come also from 
 the excessive local FUV radiation field, which dissociates CO preferentially, 
and from the small size of the clouds in the outer regions of some galaxies. 
M83 shows particularly strong FUV emission in its outer regions, with respect to others galaxies:
the FUV flux ranges between 0.2 and 1.2 $\times$ 10$^{-2}$ MJy/sr over our region.
The resolution of the FUV observations is 100 pc, at the distance of M83, so we
cannot know exactly the radiation felt by each cloud. Although the FUV flux
is much smaller than in the inner galaxy disk, clouds are smaller
and less numerous in outer parts of galaxies (here at half metallicity), which  
explains the more extended CO dissociation. 
This could be explored further through follow-up observations of the dust continuum in the Rayleigh-Jeans
domain in the same region, another independent tracer of the molecular gas.

\section{Conclusions}

We have reported about a mosaic of CO(2-1) observations obtained with ALMA in the M83 outer disk, 
rich in atomic gas and UV emission.
M83 is the prototype of spiral galaxies with an extended XUV disk. 

Our aims were to map CO emission in a small region $r_{gal} =11 kpc$ from the galaxy center. The result is a dearth
of CO emission, leading to the following conclusions: 

  \begin{enumerate}
 \item An automatic search of CO emission in the data cube provides tentative detections
of 14 clouds, but the CO(2-1) signal at 4-5 $\sigma$ does not correspond spatially and spectrally to any other
tracer. They are therefore considered as false detections.

 \item  Reversely, we searched the CO map and extract the spectra at the peak of the star formation regions
traced by H$\alpha$. This did not provide  any CO detection higher than 3$\sigma$.

 \item We have stacked all the CO pixels, at the places where significant HI signal
is found, shifting their velocity scale to a common central velocity,
expected from the HI signal. This gives a hint of detection with a profile 
width of $\Delta V$ = 14 km/s. The corresponding H$_2$ mass all over
the 4 x 2 kpc area is only 2 x 10$^6$ M$_\odot$.  The H$_2$-to-HI mass ratio over
this region is $<$ 3 $\times$ 10$^{-2}$.

 \item We display the CO upper limits towards the star forming regions
in the Kennicutt-Schmidt diagram, and the depletion time
to consume the molecular gas is lower ($<$ 3 x 10$^8$ yr) 
than in normal galaxy disks (3 x 10$^9$ yr). We could have expected to 
find in some pixels of 17 x 13pc GMC masses of 10$^6$ M$_\odot$,
while the 3$\sigma$ upper limits are  $\sim$  10$^4$ M$_\odot$.

 \item The explanation for this lack of CO emission could be due partly to a 
low metallicity, since the gas abundance in this region
of M83 is half solar. The average metallicity of the H$_{II}$ regions is $12+log(O/H)=8.4$ \cite{bresolin09}.
 Other causes of the dearth in CO may be the strong UV field, and low global 
density of gas and dust;
 the gas is predominantly H$_2$ but most carbon is not in CO molecules.
In the photodissociation regions, the carbon is in C and C+.
  
 \item In the outer parts of galaxies, at low gas surface density, the size of the clouds are likely to be smaller
than in the disk, and less self-shielded. The CO column density in each cloud is then not sufficient to
avoid dissociation, and the region is dominated by C+ emission.

 \end{enumerate} 

\begin{acknowledgements}
 We are very grateful to the referee for very useful
comments to improve and clarify the paper.
ICB would like to acknowledge the financial support from CAPES during this project.
    The ALMA staff in Chile and ARC-people at IRAM are gratefully acknowledged for their
help in the data reduction.
This paper makes use of the following ALMA data: ADS/JAO.ALMA\#2013.1.00861.S.
ALMA is a partnership of ESO (representing its member states), NSF (USA) and NINS (Japan),
together with NRC (Canada) and NSC and ASIAA (Taiwan), in cooperation with the Republic of
Chile. The Joint ALMA Observatory is operated by ESO, AUI/NRAO and NAOJ.
The National Radio Astronomy Observatory is a facility of the National Science Foundation
operated under cooperative agreement by Associated Universities, Inc.M.R. wishes to acknowledge support from CONICYT(CHILE) through FONDECYT grant 
No1140839 and partial support from project BASAL PFB-06.
\end{acknowledgements}

%--------------------------------------------------------------------

% WARNING
%-------------------------------------------------------------------
% Please note that we have included the references to the file aa.dem in
% order to compile it, but we ask you to:
%
% - use BibTeX with the regular commands:
%   \bibliographystyle{aa} % style aa.bst
%   \bibliography{Yourfile} % your references Yourfile.bib
%
% - join the .bib files when you upload your source files
%-------------------------------------------------------------------

\bibliographystyle{aa}
\bibliography{m83_bib_r}
%-------------------------------------------------------------------

\end{document}